\documentstyle[epsfig]{mn}

\newif\ifAMStwofonts

\ifoldfss
  \newcommand{\rmn}[1] {{\rm #1}}

  \ifCUPmtlplainloaded \else
    \NewTextAlphabet{textbfit} {cmbxti10} {}
    \NewTextAlphabet{textbfss} {cmssbx10} {}
    \NewMathAlphabet{mathbfit} {cmbxti10} {} 
    \NewMathAlphabet{mathbfss} {cmssbx10} {} 
  \fi
  \ifAMStwofonts
  \ifCUPmtlplainloaded \else
      \NewSymbolFont{upmath} {eurm10}
      \NewSymbolFont{AMSa} {msam10}
      \NewMathSymbol{\upi}     {0}{upmath}{19}
      \NewMathSymbol{\umu}     {0}{upmath}{16}
      \NewMathSymbol{\upartial}{0}{upmath}{40}
      \NewMathSymbol{\leqslant}{3}{AMSa}{36}
      \NewMathSymbol{\geqslant}{3}{AMSa}{3E}

      \let\leq=\leqslant \let\le=\leqslant
       \let\ge=\geqslant
    \fi
  \fi
\fi 

\ifnfssone
  \newmathalphabet{\mathit}
  \addtoversion{normal}{\mathit}{cmr}{m}{it}
  \addtoversion{bold}{\mathit}{cmr}{bx}{it}
  \newcommand{\rmn}[1] {\mathrm{#1}}

  \newmathalphabet{\mathbfit} 
  \addtoversion{normal}{\mathbfit}{cmr}{bx}{it}
  \addtoversion{bold}{\mathbfit}{cmr}{bx}{it}
  \newmathalphabet{\mathbfss} 
  \addtoversion{normal}{\mathbfss}{cmss}{bx}{n}
  \addtoversion{bold}{\mathbfss}{cmss}{bx}{n}
  \ifAMStwofonts
    \ifCUPmtlplainloaded \else
      \UseAMStwoboldmath
      \makeatletter
      \new@mathgroup\upmath@group
      \define@mathgroup\mv@normal\upmath@group{eur}{m}{n}
      \define@mathgroup\mv@bold\upmath@group{eur}{b}{n}
      \edef\UPM{\hexnumber\upmath@group}
      \new@mathgroup\amsa@group
      \define@mathgroup\mv@normal\amsa@group{msa}{m}{n}
      \define@mathgroup\mv@bold\amsa@group{msa}{m}{n}
      \edef\AMSa{\hexnumber\amsa@group}
      \makeatother
      \mathchardef\upi="0\UPM19
      \mathchardef\umu="0\UPM16
      \mathchardef\upartial="0\UPM40
      \mathchardef\leqslant="3\AMSa36
      \mathchardef\geqslant="3\AMSa3E

      \let\leq=\leqslant \let\le=\leqslant
       \let\ge=\geqslant
    \fi
  \fi
\fi 

\ifnfsstwo
  \newcommand{\rmn}[1] {\mathrm{#1}}

  \DeclareMathAlphabet{\mathbfit}{OT1}{cmr}{bx}{it}
  \SetMathAlphabet\mathbfit{bold}{OT1}{cmr}{bx}{it}
  \DeclareMathAlphabet{\mathbfss}{OT1}{cmss}{bx}{n}
  \SetMathAlphabet\mathbfss{bold}{OT1}{cmss}{bx}{n}
  \ifAMStwofonts
    \ifCUPmtlplainloaded \else
      \DeclareSymbolFont{UPM}{U}{eur}{m}{n}
      \SetSymbolFont{UPM}{bold}{U}{eur}{b}{n}
      \DeclareSymbolFont{AMSa}{U}{msa}{m}{n}
      \DeclareMathSymbol{\upi}{0}{UPM}{"19}
      \DeclareMathSymbol{\umu}{0}{UPM}{"16}
      \DeclareMathSymbol{\upartial}{0}{UPM}{"40}
      \DeclareMathSymbol{\leqslant}{3}{AMSa}{"36}
      \DeclareMathSymbol{\geqslant}{3}{AMSa}{"3E}

      \let\leq=\leqslant \let\le=\leqslant
       \let\ge=\geqslant
    \fi
  \fi
\fi 

\ifCUPmtlplainloaded \else
  \ifAMStwofonts \else 
    \def\upi{\pi}
    \def\umu{\mu}
    \def\upartial{\partial}
  \fi
\fi

\title{Power Spectrum Analysis of the ESP Galaxy Redshift Survey}
\author[E. Carretti et al.]
       {E.~Carretti,$^{1,2}${\thanks{carretti@tesre.bo.cnr.it}}  
        C.~Bertoni,$^3${\thanks{bertoni@astbo1.bo.cnr.it}}
	A.~Messina,$^4${\thanks{messina@cs.unibo.it}}
	E.~Zucca$^5${\thanks{zucca@bo.astro.it}} and
        L.~Guzzo$^6${\thanks{guzzo@merate.mi.astro.it}}\\
        $^1$Istituto Te.S.R.E., Via Gobetti 101, I-40129 Bologna, ITALY\\
        $^2$Dipartimento di Astronomia, Via Ranzani 1, I-40127 Bologna, ITALY\\
        $^3$Istituto di Radioastronomia, Via Gobetti 101, I-40129 Bologna, ITALY\\
        $^4$Dipartimento di Scienze dell'informazione, Via Mura Anteo Zamboni 7,
            I-40126 Bologna, ITALY\\
        $^5$Osservatorio Astronomico di Bologna, Via Ranzani 1,
            I-40127 Bologna, ITALY\\
        $^6$Osservatorio Astronomico di Brera, Via Bianchi 46, I-23807,
            Merate(LC), ITALY\\}
\date{Accepted.
      Received;
      in original form}

\pagerange{\pageref{firstpage}--\pageref{lastpage}}
\pubyear{0000}

\begin{document}

\maketitle

\label{firstpage}

\begin{abstract}
We measure the power spectrum of the galaxy distribution in the ESO Slice
Project (ESP) galaxy redshift survey.  We develope a technique to describe the
survey window function analytically, and then deconvolve it from the
measured power spectrum using a variant of the Lucy method.  We test the whole
deconvolution procedure on ESP mock catalogues drawn from large N--body
simulations, and find that it is reliable for recovering the correct amplitude
and shape of $P(k)$ at $k> 0.065\, h$ Mpc$^{-1}$.  In general, the technique is
applicable to any survey composed by a collection of circular fields with
arbitrary pattern on the sky, as typical of surveys based on fibre
spectrographs.   The estimated power spectrum has a well--defined power--law
shape $k^n$ with $n\simeq -2.2$ for $k\ge 0.2\,h$ Mpc$^{-1}$, and a smooth bend
to a flatter shape ($n\simeq -1.6$) for smaller $k$'s. 
The smallest wavenumber, where a meaningful reconstruction can be performed
($k\sim 0.06\,h$~Mpc$^{-1}$),
does not allow us to explore the range of
scales where other power spectra seem to show a flattening and hints for a
turnover.  We also find, by direct comparison of the Fourier transforms, that the
estimate of the two--point correlation function $\xi(s)$ is much less sensitive
to the effect of a problematic window function as that of the ESP, than the
power spectrum.   Comparison to other surveys shows an excellent agreement with
estimates from blue--selected surveys.  In particular, the ESP power
spectrum is virtually indistinguishable from that of the Durham-UKST survey over
the common range of $k$'s, an indirect confirmation of the quality of the
deconvolution technique applied.  

\end{abstract}


\section{Introduction}
The quantitative characterisation of the galaxy distribution is a
major aim in the study of the large-scale structure of the Universe.
During the last 20 years, several surveys of galaxy redshifts have shown that
galaxies are grouped in clusters and superclusters, drawing structures surrounding
large voids (see e.g. Guzzo 1999 for a review).  The 
power spectrum of the galaxy distribution provides a 
concise
statistical description of the observed clustering that, 
under some assumptions on its relation to the 
mass distribution,
represents an important test for different structure formation scenarios
(e.g. Peacock 1997 and references
therein).  Indeed, under the assumption of Gaussian fluctuations, the
power spectrum totally describes the statistical properties of the
matter density field (e.g. Peebles 1980).  

In recent years, several estimates of the galaxy power spectrum have
been obtained  
using galaxy samples selected at different wavelenghts: radio (Peacock
\& Nicholson 1991), infrared (Feldman, Kaiser \& Peacock 1994, Fisher
et al. 1993, Sutherland et al. 1999) and optical (Park et al. 1994,
da~Costa et al. 1994, Tadros \& Efstathiou 1996, Lin et al. 1996,
Hoyle et al. 1999), to mention the most recent ones.

The ESO Slice Project redshift survey (ESP, Vettolani et al. 1997,
1998) is one of the two deepest wide--angle surveys currently
available, inferior only to the larger Las Campanas Redshift Survey (LCRS,
Shectman et al. 1996).  During the last few years, it has produced a number of
statistical results on the properties of optically--selected galaxies, as
e.g. the luminosity function (Zucca et al. 1997) or the correlation function
(Guzzo et al. 2000).   The geometry of the 
survey (a thin row of
circular fields, resulting in an essentially 2D slice in space) is such
that an estimate of the power spectrum 
represents a
true challenge.  
In this paper we present the results of a detailed analysis that
overcomes these difficulties, 
producing a reliable measure of the power
spectrum from the ESP redshift data.  The technique developed here to cope with
the specific geometry of the survey is potentially interesting also for
application to other surveys consisting of separate patches on the 
sky, as could be the case, for example, of preliminary sub--samples of the 
ongoing SDSS (Margon 1998) and 2dF (Colless 1998) surveys. 

The outline of the paper is as follows. We shall first recall the main
features of the ESP survey and the sample selection
(section~\ref{ESP_sur}), then discuss the power spectrum estimator
adopted for the analysis (section~\ref{ESP_est}) and the numerical
tests performed in order to check its validity range (section~\ref{ESP_num}).
We shall then present the estimated power spectrum
(section~\ref{ESP_pow}) and its consistency with the correlation
function (section~\ref{ESP_corr}), and then discuss it in comparison 
to the results from other surveys (section~\ref{ESP_comp}).    
Section~\ref{conc} summarises the results obtained, drawing some conclusions. 

\section{The ESO Slice Project}\label{ESP_sur}

\begin{figure*}
\epsfxsize=\hsize \epsfbox{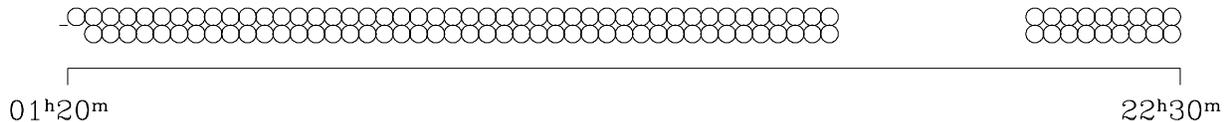} \hfil
  \caption{The area covered by the ESP survey on the sky 
           consists of a set of 107 circular
           fields of $16'$ radius.  
           As shown in this figure, they are arranged 
           into 2 parallel rows and draw two thin slices
	   over the celestial sphere
           of about $22\degr \times 1\degr$ and
	   $5\degr \times 1\degr$ respectively,
           separated by $\sim 5\degr$ (from Vettolani et al. 1997).} 
\label{fig1}
\end{figure*}

\begin{figure*}
  \centerline{
    \psfig{figure=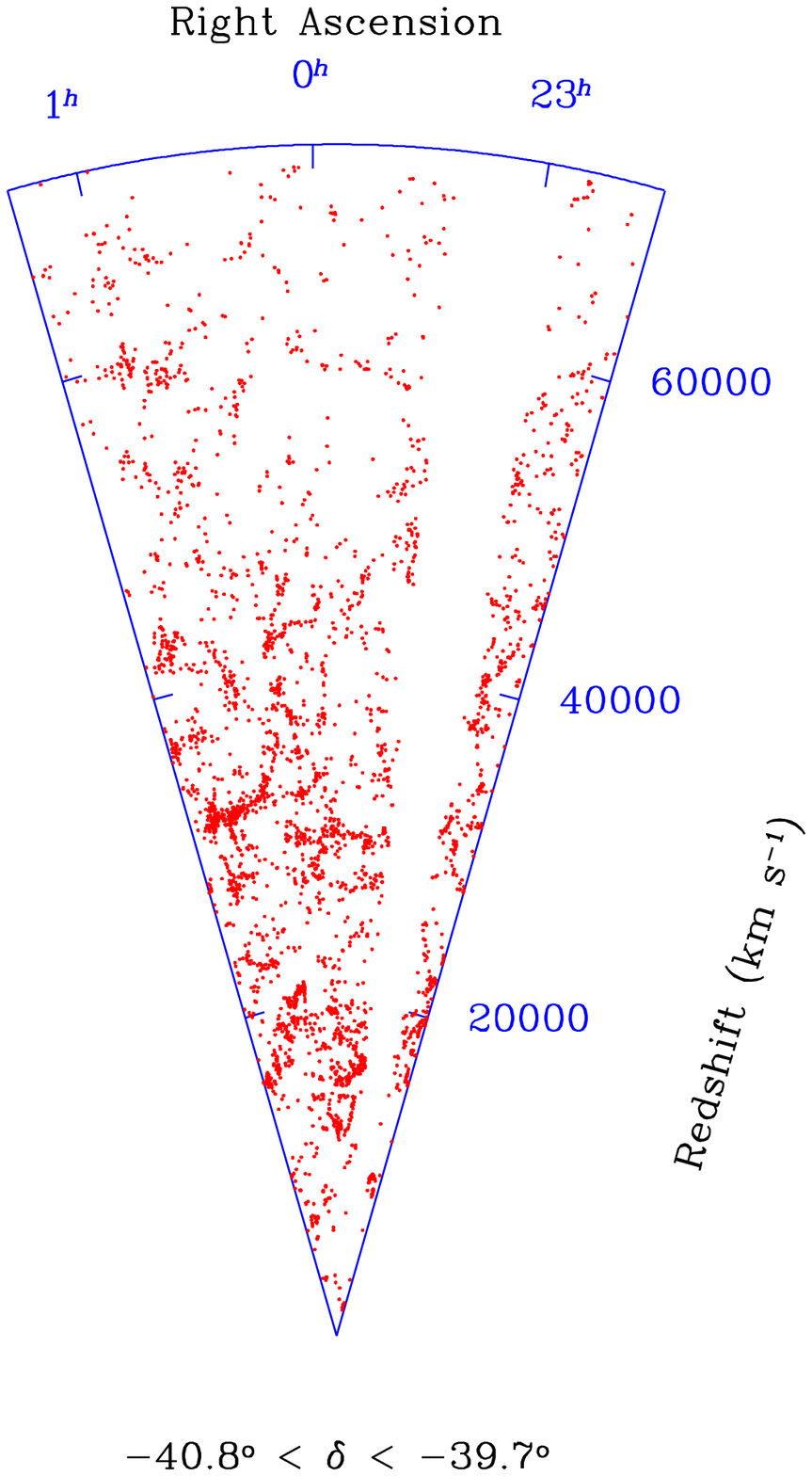, width=1.0\hsize, angle=0}
  }
  \caption{The galaxy distribution in the ESP redshift survey.} 
\label{fig2}
\end{figure*}

The ESO Slice Project galaxy redshift survey (ESP, Vettolani et
al. 1997, 1998) was constructed between 1993 and 1996 to fill the 
gap that existed at the time between shallow, wide angle surveys as the CfA2, and
very deep, one-dimensional pencil beams.  The survey was designed in order to
allow the sampling of volumes larger than the maximum sizes of known structures
and an unbiased estimate of the luminosity function of field galaxies to very
faint absolute magnitudes.  The survey and the data catalogue are described in
detail in Vettolani et al. (1997, 1998).  Here we limit ourselves to a summary
of the main features relevant for the present analysis.
\\
The ESP survey (see Figures~\ref{fig1} and~\ref{fig2})
extends over a strip of $\alpha \times \delta = 22\degr \times 
1^{\rm o}$, plus a nearby area of $5\degr \times 1\degr$,
five degrees west of the main 
strip, in the South Galactic Pole region ($ 22^{h} 30^m \le \alpha \le
01^{h} 20^m $, at a mean  
declination of $ -40^{\rm o} 15'$ (1950)). This region was covered with a regular 
grid of adjacent circular fields, with a diameter of 32 arcmin each, 
corresponding to the field of view of the multifibre spectrograph OPTOPUS 
(Avila et al. 1989) at the ESO 3.6~m telescope. The total solid angle
covered by the survey is 23.2 square degrees and its position on the
sky was chosen in order to minimize galactic absorption ($-75^{\rm o} \la
b^{II} \la -60^{\rm o}$).  
The target objects, with a limiting magnitude $b_J \leq 19.4$, were selected 
from the Edinburgh--Durham Southern Galaxy Catalogue (EDSGC, Heydon--Dumbleton 
et al. 1989).
A total of 4044 objects were observed, corresponding to $\sim 90\%$ of the 
parent photometric sample and selected to be a random subset of the
total catalogue with respect to both magnitude and surface brightness.
The total number of confirmed galaxies with reliable redshift measurement
is 3342, while 493 objects turned out to be stars and 1 object is a quasar
at redshift $z \sim 1.174$. No redshift measurement could be obtained for the
remaining 208 spectra.  
As discussed in Vettolani et al. (1998), the magnitude distribution of
the missed galaxies is consistent with a random extraction of the
parent population.
About half of the ESP galaxies present spectra with emission lines.
Particular attention was paid to the redshift quality and several checks 
were applied to the data, using 1)~multiple observations of $\sim 200$ 
galaxies, 2)~$\sim 750$ galaxies for which the redshift from both absorption 
and emission line is available (Vettolani et al. 1998, Cappi et al. 1998). 
More 
details about the data reduction and sample completeness are
reported in Vettolani et al. (1997, 1998).  

Given the magnitude--limited nature of the survey, the computation of
a clustering statistics like the power spectrum requires the knowledge
of the selection function.  This is defined as the expected probability to
detect a galaxy at a redshift $z$ and can be expressed as 
\begin {equation}
        s(z) = {\int_{{\rm max}[L_1,L_{\rm min}(z)]}^{+\infty} \phi(L) dL \over
                     \int_{L_1}^{+\infty} \phi(L) dL},
\end {equation}
where $\phi(L)$ is the luminosity function, $L_1$ is the minimum
luminosity of the sample and $L_{\rm min}(z)$ is the minimum luminosity detectable
at redshift $z$, given the sample limiting magnitude. 

In the ESP survey the minimum luminosity corresponds to an absolute magnitude
$M_{b_J,1}=-12.4+5\,\log h$ ($h$ is the Hubble constant in units of $100\, 
{\rmn km\, s}^{-1}{\rmn Mpc}^{-1}$). $L_{\rm min}(z)$ is the luminosity of a galaxy
at redshift $z$ with an apparent magnitude equal to the apparent magnitude
limit $b_J = 19.4$. The corresponding absolute magnitude is given by
\begin {equation}
        b_J-M_{b_J} = 25 + 5\,\log D_L(z) + K(z),
\end {equation}
where $D_L$ is the luminosity distance in Mpc and $K(z)$ is the K-correction. 
$D_L$ is given by the Mattig formula (1958), which depends on the assumed
cosmological model. For all ESP computations we assume a flat universe
with $\Omega_{\rm o} = 1$ and $\Lambda=0$.   Before proceeding to the computation of
luminosity distances, we have converted the observed heliocentric redshifts in
the catalogue to the Cosmic Microwave Background (CMB) rest frame using a
standard procedure, as described in Carretti (1999).
The luminosity distance is then given by
\begin {equation}
        D_L(z) = {2 c\over H_{\rm o}} (1+z) \left(1- {1\over \sqrt{1+z}}\right) 
\end {equation}

The K-correction is a function of 
redshift 
and 
morphological type,
but the latter is not directly available for ESP galaxies.  
Following Zucca et al. (1997), we use an average K-correction,
weighted over the expected morphological mixture at each $z$.  See Zucca et
al. (1997, cfr. their figure 1) for the details of this computation. 
A recent principal component analysis of the spectra (Scaramella,
priv. comm.) confirms the adequacy of this mean correction.  

The luminosity function is such that $\phi(L)dL$ gives the  density of 
galaxies with luminosity $L\in[L,L+dL[$. The ESP luminosity function
is well approximated by a Schechter (1976) function (Zucca et
al. 1997)
\begin {equation}
        \phi(L)dL = \phi^* \left({L\over L^*}\right)^{\alpha} e^{-L/L^*}
                   d\left({L\over L^*}\right)
\end {equation}
with best fit parameters $\alpha = -1.22$,
$M_{b_J}^* = -19.61 + 5\log h$ and $\phi^* = 0.020 h^3$ Mpc$^{-3}$.
In reality, 
as shown by Zucca et al. (1997),
for $M_{b_J} > -16 + 5\log h$ the faint end 
steepens
with respect to the Schechter 
form and 
the overall shape 
is better described by adding an
extra power law.  Nevertheless, this is relevant only for the
very local part of the sample and a description of the selection
function using a simple Schechter fit is perfectly adequate for our purposes.

Another quantity to be taken into account 
for clustering analyses
is the redshift completeness 
of the 107 fields, as not all target galaxies at the photometric limit were
succesfully measured.
\begin{figure*}
 \epsfxsize=\hsize \epsfbox{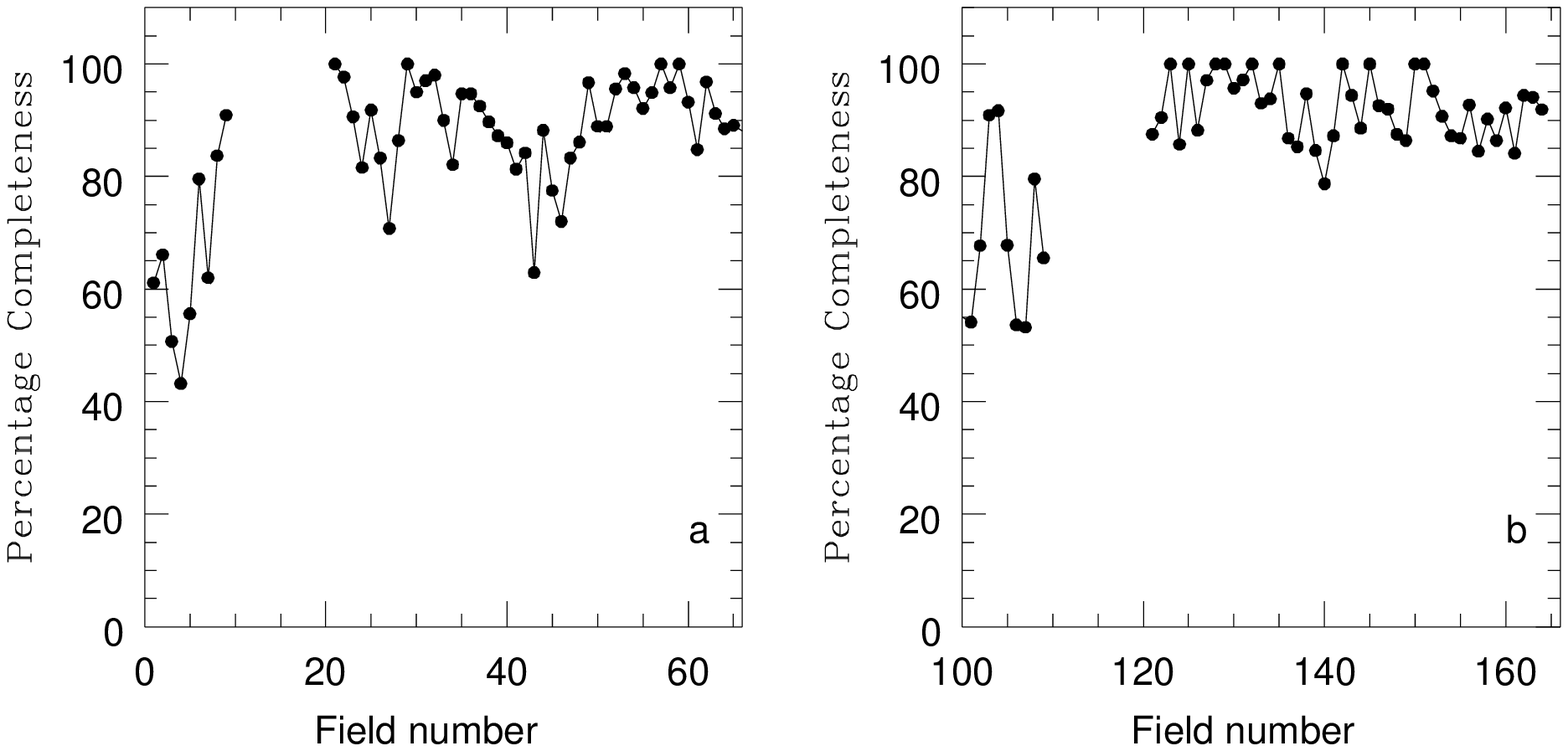} \hfil
 \caption{The redshift completeness within the ESP fields, i.e. the
fraction of galaxies with measured redshift with respect to the total
number of galaxies in the photometric sample in each field. Field
numbers are as reported in the catalogue (Vettolani et al., 1998).
The two panels are for the fields in the northern (a) and southern
 (b) rows respectively.}
 \label{fig3}
\end{figure*}
This can be expressed as (Vettolani et al. 1998)
\begin {equation}
        C = { N_Z \over N_T - N_S - 0.122N_{NO}},
\end {equation}
where, for each field, $N_T$ is the total number of objects in the
photometric catalogue, $N_Z$ is the number of reliable galaxy redshifts,
$N_{NO}$ is the number of not observed objects, $N_S$ is the number of
stars and 0.122 is the fraction of stars in the spectroscopic sample. 
In Figure~\ref{fig3} we plot the completeness values for each field. 
Field numbers $<100$ denote fields in the northern row, while the
others refer to the southern one.  

The power spectrum analysis has been performed on both 
volume--limited and magnitude--limited 
subsamples of the survey.
Volume--limited samples include all galaxies intrinsically more
luminous than a given absolute magnitude $M_{\rm lim}$ and within the
maximum redshift $z_{\rm max}$ at which such magnitude can still be
detected within the survey apparent magnitude limit.  
In such a case, the expected mean density of galaxies does not
vary with distance. Magnitude--limited catalogues, by definition, are
simply subsets of all galaxies in the survey to a given apparent
magnitude, possibly with the addition of an upper distance cut 
$z_{\rm max}$
above which 
the value of the selection function becomes
too small.
Magnitude--limited 
samples 
contain more objects, 
but the mean ensemble properties (as e.g. the galaxy 
luminosity distribution) vary with distance.  
We extract 
from the ESP survey
two magnitude--limited samples with different $z_{\rm max}$ limit, plus
one volume--limited sample
with 
$M_{\rm lim} \le -20.1 + 5 \log h$. (For simplicity, we shall omit
hereafter the $5 \log h$ term).

For the estimate of the power spectrum, comoving distances are computed
for each galaxy as $D_c(z) = D_L(z)/(1+z) $.
The 
uncertainty 
introduced in $D_c$
because of 
our ignorance of the correct cosmological model 
amounts to less than 5\% 
for
a typical redshift $z=0.20$, when the value of $\Omega_{\rm o}$ 
is changed from 1 to 0.2.

\begin{table}
 \centering
  \caption{Parameters of the samples extracted from the ESP survey:
           $z_{\rm max}$ is the maximum redshift, $D_{\rm max}$ the maximum comoving distance
           in $h^{-1}$ Mpc unit, $M_{\rm lim}$ the absolute magnitude limit for
           the volume--limited sample (we omit the $5 \log h$ term)
           and $N$ is the galaxy number.}
  \begin{tabular}{@{}lrccr@{}}
     \hline \\
     Sample & $z_{\rm max}$ & $D_{\rm max}$ & $M_{\rm lim}$ & $N$ \\
            &           & ($h^{-1}$~Mpc)&   &   \\
     \hline\\
     ESPm523 & $0.20$ & $523$ &  & $3092$ \\
     ESPm633 & $0.25$ & $633$ &  & $3306$ \\
     ESP523 & $0.20$ & $523$ & $-20.1$  & $481$ 
\end{tabular}
\label{tab_sam}
\end{table}

\section[]{Power Spectrum Estimator}\label{ESP_est}
The galaxy power spectrum can be defined as
\begin{equation}
  P(k) = \int \xi(x) e^{-i{\bf k}\cdot {\bf x}}\, d{\bf x},
        \label{Pkeq}
\end{equation}
where $\xi(x)$ is the two--point correlation function,
${\bf x}$ and ${\bf k}$ are the comoving position and
wavenumber vectors respectively, while $x = |{\bf x}|$ and $k = |{\bf k}|$.
Under the hypothesis of homogeneity and isotropy, $P(k)$ and $\xi(x)$ are
functions only of $k$ and $x$ respectively.
By definition the two--point correlation function
can be also written
\begin{equation}
  P(k) \propto |\hat{\delta}(k)|^2 ,
        \label{Pkeqdef}
\end{equation}
where $\hat{\delta}({\bf k})$ is the Fourier transform of 
the density contrast of the galaxies.

In this paper we follow the Fourier notation
\begin{equation}
      \hat{f}({\bf k}) = \int f({\bf x}) e^{-i{\bf k}\cdot {\bf x}}\, d{\bf x},
\end{equation}
\begin{equation}
       {f}({\bf x}) = {1\over (2\pi)^3}\int \hat{f}({\bf k})
                      e^{i{\bf k}\cdot {\bf x}}\, d{\bf k}.
\end{equation}

To compute the power spectrum of
galaxy density fluctuations from the observed galaxy distribution,
we use a traditional Fourier method (cfr. Carretti 1999 for details),
as developed by several authors (e. g. cfr.
Peebles 1980, Fisher et al. 1993, Feldman et al. 1994, Park et al. 1994,
Lin et al. 1996). We also apply a correction
(Tegmark et al. 1998), that accounts for 
our ignorance on the true value of the mean density of galaxies
(Peacock \& Nicholson 1991).

Given a sample of $N$ galaxies of positions ${\bf x}_j$ and weights
$w_j$,
an estimate of the Fourier transform of density contrast 
is given by
\begin{equation}
        \hat{\tilde{\delta}}({\bf k}) = {V \over \sum_{j=1}^N w_j}
            \sum_{j=1}^N w_j e^{-i{\bf k}\cdot{\bf x}_j} - \hat{W}({\bf k}),
        \label{deltilkeq}
\end{equation}
where $V$ is the volume of the sample and 
$\hat{W}({\bf k})$ is the Fourier transform of the survey
window function (hereafter a $\sim$ will denote the quantities estimated
from the data). 
The window function ${W}({\bf x})$ is 1 within the volume covered by the
sample and 0 elsewhere, so it can be described as an ensemble of 107 cones.
This geometry allows us to obtain analytically the Fourier transform
$\hat{W}({\bf k})$ as the sum of the Fourier transform of each cone.
In equation~\ref{deltilkeq}
each galaxy contributes with some weight $w_j$. In a
volume--limited catalogue all galaxies have equal weight, i.e.
$w_j \equiv 1$.
In a magnitude--limited catalogue the expected galaxy density decreases
with the distance 
according to 
the selection function. Thus, the 
simplest form for the weight for a galaxy
is given by the inverse of the selection function at its redshift $z_j$
\begin{equation}
        w_j = {1\over s({z_j})}.
\end{equation}
If the catalogue completeness is $C<1$, the previous weight should be
modified as
\begin{equation}
        w_j = {1\over C({\bf x}_j)}
\end{equation}
for volume--limited catalogues, and as
\begin{equation}
        w_j = {1\over s(z_j) C({\bf x}_j)},
\end{equation}
for magnitude--limited catalogues. $C({\bf x}_j)$ 
is the completeness of the sample at the position of the $j^{th}$ galaxy.

Our adopted power spectrum estimator is defined with respect to
$\hat{\tilde{\delta}}({\bf k})$ by the following equation
(Tegmark et al. 1998)
\begin{equation}
        \tilde{{P_c}}({\bf k}) =
               {| \hat{\tilde{\delta}}({\bf k})|^2 - \tilde{b}({\bf k})
                                   \over A({\bf k})},
    \label{Pconveq}
\end{equation}
where
\begin{equation}
        A({\bf k}) = \left(
                       1 - \left|{\hat{W}({\bf k})\over
                                  \hat{W}({\bf 0})}\right|^2
                     \right) V
        \label{Akleq}
\end{equation}
accounts for our ignorance of the mean galaxy density, while
\begin{equation}
    \tilde{b}({\bf k}) = {V^2 \over \left(\sum_{j=1}^N w_j \right)^2}
                           \sum_{j=1}^N w_j^2
                           \left| e^{-i{\bf k}\cdot{\bf x}}
                           - {\hat{W}({\bf k})\over
                           \hat{W}({\bf 0})}\right|^2
\end{equation}
is the shot noise
correction due to the finite size of the sample.

The observed power spectrum estimated by equation~\ref{Pconveq}  is
related to the true power spectrum $P(k)$ by
\begin{equation}
  \tilde{{P_c}}({\bf k}) = {1\over (2\pi)^3 A({\bf k})} \int P(k')
			    \phi({\bf k}, {\bf k}') d{\bf k}',
\end{equation}
where
\begin{equation}
  \phi({\bf k}, {\bf k}') = \left|\hat{W}({\bf k} - {\bf k}')
                         - {\hat{W}({\bf k})\over \hat{W}({\bf 0})}
                        \hat{W}(-{\bf k}')\right|^2 .
\end{equation}
For wavenumbers ${\bf k}$ such that $|\hat{W}({\bf k})| \ll |\hat{W}({\bf 0})|$
this equation reduces to the convolution between $P(k)$ and
$|\hat{W}({\bf k})|^2$.

To describe the convolved power spectrum we choose to average
$\tilde{{P_c}}({\bf k})$ over all directions
\begin{eqnarray}
     \tilde{{P_c}}(k) 
        &=& \left<\tilde{{P_c}}({\bf k})\right> \nonumber \\
        &=& {1\over 4\pi } \int_{{\bf \Omega}_{k}}
            \tilde{{P_c}}({\bf k}) d{\bf \Omega}_{k} \nonumber \\
        &=& \int_0^{\infty} {k'}^2P(k')\chi(k,k')\, dk',
    \label{conveq}
\end{eqnarray}
where ${{\bf \Omega}_{k}}$ is the sphere defined by wavenumbers of amplitude
$k$. The kernel of this integral equation is given by 
\begin{equation}
   \chi(k,k') = {1\over 2(2\pi)^4 V}
                  \int_{{\bf \Omega}_{k}} \int_{{\bf \Omega}_{k'}}
	    	  \psi({\bf k}, {\bf k}')\, 
                  d{\bf \Omega}_{k} d{\bf \Omega}_{k'},
   \label{conv1eq}
\end{equation}
where
\begin{equation}
   \psi({\bf k}, {\bf k}') = {
                   \left|\hat{W}({\bf k} - {\bf k}') - 
		   \hat{W}(-{\bf k}') \; \hat{W}({\bf k}) /
                   \hat{W}({\bf 0}) \right|^2
                   \over
                   1 - \left|\hat{W}({\bf k}) /
                    \hat{W}({\bf 0})\right|^2
                   }
   \label{conv1beq}
\end{equation}
and
${\bf \Omega}_{k'}$ is the sphere defined by wavenumbers of amplitude $k'$.

The Fourier transform of the window function has been analytically 
computed as the sum of the Fourier transforms of all
the 107 cones. In reality, the cones are slightly overlapped but the
small common volume (2.85\%) allows us to make the
assumption of disjoined cones.

The small width of one ESP cone allows us
to analytically compute its Fourier transform.
Let $r_{\rm o}$ be the cone height and $\Delta\theta \ll 1$~rad its width
($\Delta\theta = 16' = 0.00465$~rad).
In the simple case of a cone centered on the $z$ axis,
the Fourier transform is
\begin{equation}
   \hat{W}_c ({\bf k}) = \int_0^{r_{\rm o}} dr\, r^2 \int_0^{2\pi} d\phi  
                 \int_0^{\Delta\theta} d\theta\,\sin\theta\, e^{-i{\bf k}\cdot {\bf r}}.
\end{equation}
Taking into account the small value of $\Delta\theta$,
the integrand can be approximated to first order
in $\theta$, resulting in
\begin{eqnarray}
   \hat{W}_c ({\bf k}) &=& \int_0^{r_{\rm o}} dr\, r^2 \int_0^{2\pi} d\phi  
                           \int_0^{\Delta\theta} d\theta\,\sin\theta\, e^{-ikr\cos\alpha} \nonumber \\
                       &=& 2\pi [1 - \cos(\Delta\theta)] \int_0^{r_{\rm o}} dr\,
		           r^2 e^{-ikr\cos\alpha}.
    \label{wck1eq}
\end{eqnarray}
This expression depends on $r_{\rm o}$, $\Delta\theta$ and $\cos\alpha$, where $\alpha$ is
the angle between ${\bf k}$ and the $z$ axis.
For a generic cone along the direction
$(\theta_{\rm o}, \phi_{\rm o})$ a rotation can be applied 
in order to bring the cone
on the $z$ axis. So, the Fourier transform is
\begin{eqnarray}
  \hat{W}_c ({\bf k}) &=& \int_0^{r_{\rm o}} dr\, r^2 \int_{{\bf \Omega}_{(\theta_{\rm o},\phi_{\rm o})}} d{\bf \Omega}  
                          e^{-i{\bf k}\cdot {\bf r}} \nonumber \\
                      &\approx& 2\pi [1 - \cos(\Delta\theta)] \int_0^{r_{\rm o}} dr\, r^2 e^{-ikr\cos\gamma}, 
   \label{wck1beq}
\end{eqnarray}
where ${\bf \Omega}_{(\theta_{\rm o},\phi_{\rm o})}$
is the solid angle of the cone
centered on $(\theta_{\rm o},\phi_{\rm o})$, and
$\gamma$ is the angle between the wave number
${\bf k}$ and the direction $(\theta_{\rm o},\phi_{\rm o})$. 
By solving the integral one gets
\begin{eqnarray}
  \hat{W}_c ({\bf k}) &=& 2\pi\,i{[1 - \cos(\Delta\theta)] 
                                  \over k\cos\gamma }\times \nonumber \\
                      & & \!\!\!\!\!\!\!\!\!\!\!\!\!\!\!\!\!\!\!\!\!\!\!\!\!\!\!\!\!\!\!
		          \left[
                          r_{\rm o}^2 e^{-ikr_{\rm o}\cos\gamma}
                          - i {2 r_{\rm o} e^{-ikr_{\rm o}\cos\gamma}\over k\cos\gamma  }
                          -2 {e^{-ikr_{\rm o}\cos\gamma} - 1 \over (k\cos\gamma)^2}
                          \right].
    \label{wck2eq}
\end{eqnarray}
Finally, the Fourier transform of the whole ESP window function is 
given by the sum of
the Fourier transform of the cones normalized to the true volume of the survey
to account for the overlapping zone
\begin{equation}
  \hat{W} ({\bf k}) = {V \over 107\, V_c} \sum_{i=1}^{107} \hat{W}_{c,i}({\bf k}),
   \label{winespeq}
\end{equation}
where $i$ is the cone index, $V_c$ is the volume of one cone 
and $V$ is the true survey volume
\begin{equation}
    V = 107\, V_c (1 - \beta),
\end{equation}
which accounts for the total volume fraction $\beta = 0.0285$ lost in
the cone overlaps.

To check the reliability of our analytic approximation,
we perform a numerical
Fourier computation. We sample the survey volume on a regular grid by assigning $1$ to the
grid points inside the window function and $0$ outside. We then perform an FFT.
This numerical computation is limited by the finite size of the grid cells,
but avoids the overlapping zones and considers the true window function. 
\begin{figure}
 \epsfxsize=\hsize \epsfbox{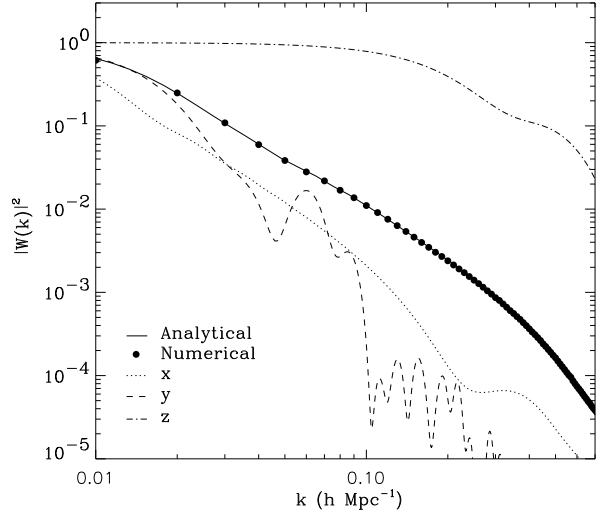} \hfil
 \caption{The power spectrum of the ESP window function (here we use this term
for the real--space survey selection window, but in other papers the
same is used for its Fourier transform directly).  The filled line is
the spherically--averaged function, computed analytically using the
machinery described in the text.  The filled points give the same
quantity, but computed numerically through a simple Montecarlo
simulation.  The figure shows also the three components of $|W(k)|^2$
in k--space.  Note how broad is the function along the direction $z$,
which has been chosen to be essentially perpendicular to the ESP main
plane, evidencing its extreme anisotropy.
}
\label{fig4}
\end{figure}
Figure~\ref{fig4} (filled points and solid line)
compares both the analytical and numerical
estimates of the window function power spectrum averaged over 
spherical shells. The difference is less than 5\%.

The strongly anisotropic geometry of the
ESP survey (see Figure~\ref{fig4})
introduces important convolution effects between the survey window
function and the galaxy distribution. 
To clean the observed power spectrum for these effects,
we have adopted Lucy's
deconvolution method (Lucy 1974; 
see also Baugh \& Efstathiou 1993 and Lin et al. 1996, for a discussion 
about its application to power spectrum estimates).

The Lucy technique is a general method to estimate the frequency
distribution $\psi(\eta)$ of a quantity $\eta$, when we know the
frequency distribution $\phi(y)$ of a second quantity $y$,
related to $\eta$ by      
\begin{equation}
   \phi(y) = \int\psi(\eta) \Pi(y|\eta)\, d\eta, 
	\label{phiinteq} 
\end{equation}
where $\Pi(y|\eta)\,dy$ is the probability that $y' \in [y,\,y\!+\!dy[$
when $\eta'=\eta$. The probability $\Pi(y|\eta)$ must be known and the
frequency distribution $\phi(y)$ is the observed one.

The solution of equation~\ref{phiinteq} can be obtained by an
iterative procedure. Let $Q(\eta|y)\,d\eta$ be the probability that
$\eta' \in [\eta,\,\eta\!+\!d\eta[$ when $y'=y$. The probability that
$y' \in [y,\,y\!+\!dy[$ and $\eta' \in [\eta,\,\eta\!+\!d\eta[$ can be
written as $\phi(y)dy\,Q(\eta|y)d\eta$ and 
$\psi(\eta)d\eta\,\Pi(y|\eta)dy$. From these two expressions and
equation~\ref{phiinteq} we obtain
\begin{equation}
    Q(\eta|y) = {\psi(\eta) \Pi(y|\eta) \over \int \psi(\eta) \Pi(y|\eta)\, d\eta},
	\label{Qdefeq}
\end{equation}
which provides the identity
\begin{equation}
    \psi(\eta) \equiv \int \phi(y)Q(\eta|y)\, dy.
    \label{psiequiveq}
\end{equation}

The latter equation cannot be solved directly, since $Q(\eta|y)$ depends on
the unknown $\psi(\eta)$ as well. Given a
fiducial model for $\psi(\eta)$ and the known probability $\Pi(y|\eta)$,
equation~\ref{Qdefeq} provides an estimate for $Q(\eta|y)$.
This and the identity~\ref{psiequiveq} allows us
to compute an
improved estimate for $\psi(\eta)$. The process can then be repeated 
until convergency.
In our specific case the equation to be solved is eq. \ref{conveq}, where 
${k'}^2\chi(k,k')$ plays the role of the probability $\Pi(y|\eta)$.
If we sample $k$ on logarithmic intervals the convolution integral
becomes

\begin{equation}
       \tilde{{P_c}}(k) = \int {k'}^3 P(k')  \chi(k,k')\ln (10)
       \,\, d(\log_{10} k')
	\label{lucylogeq}
\end{equation}
and an iterative scheme for the deconvolved spectrum can be written as
\begin{equation}
    P^{m+1}(k_i) = P^m(k_i) 
           {\sum_j {\left[\tilde{{P_c}}(k_j) / \tilde{{P_c}}^m(k_j)\right]}
            \chi(k_j,k_i)
	      \over
            \sum_{j} \chi(k_j,k_i)},
\end{equation}
where
\begin{equation}
       \tilde{{P_c}}^m(k_j) = \sum_r {k_r}^3 P^m(k_r)  
       \chi(k_j,k_r)\ln(10)\,\,\Delta
\end{equation}
and $\Delta = (\log_{10} k_{i+1} - \log_{10} k_{i})$
is the logarithmic interval, while $P^m$ denotes the $m^{th}$ estimate
of the spectrum. 

One problem with the Lucy method is that of producing a noiser and noiser
solution as the iteration converges. To avoid this, Lucy
suggests to stop the iteration after the first few steps. 
This is quite arbitrary and we prefer to follow Baugh \& Efstathiou (1993;
see also Lin et al. 1996) in applying a smoothing procedure at each step
\begin{equation}
    P^{m}(k_i) = 0.25 P^m(k_{i-1}) + 0.5 P^m(k_i) + 0.25 P^m(k_{i+1}).
\end{equation}
We use $P^0(k_i)= constant$ as initial guess for the power spectrum, but
we checked that the solution is independent of the shape of $P^0(k_i)$.
One consequence of this smoothing is that some degree of correlation
is introduced among the bins of $P(k)$.

\begin{figure}
 \epsfxsize=\hsize \epsfbox{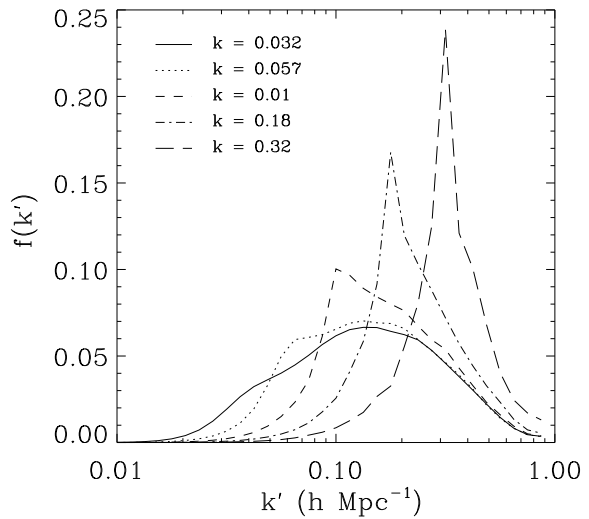} \hfil
 \caption{Behaviour of the integrand 
  of the convolution
  equation~\ref{lucylogeq} normalized with respect to
  the convolved power spectrum 
  $\tilde P(k)$ for
  some $k$ values.
  We plot the quantity
  $f(k') = {k'}^3 P(k') \chi(k,k')\ln (10)\Delta / \tilde P(k)$
  for  ($k = 0.032,\, 0.057,\, 0.1,\, 0.18,\, 0.32 h$ Mpc$^{-1}$),
  considering the kernel relative to the geometry of the $523\,h^{-1}$ Mpc
  sample.
  The power spectrum $P(k')$ of a CDM model with $\Omega = 0.4$ and
  $\Gamma = 0.2$ has been used.
 }
\label{fig5}
\end{figure}

The importance of the convolution effects on different scales
can be estimated 
by plotting the integrand 
of eq. (\ref{lucylogeq}) as a function of $k'$, for different 
values of $k$ (Figure~\ref{fig5}).
If the window was a large and 
regular sample of the Universe, the plots would be sharply peaked
at $k = k'$, as it actually happens for large values of $k$ (small
scales).
\begin{figure*}
 \epsfxsize=0.9\hsize \epsfbox{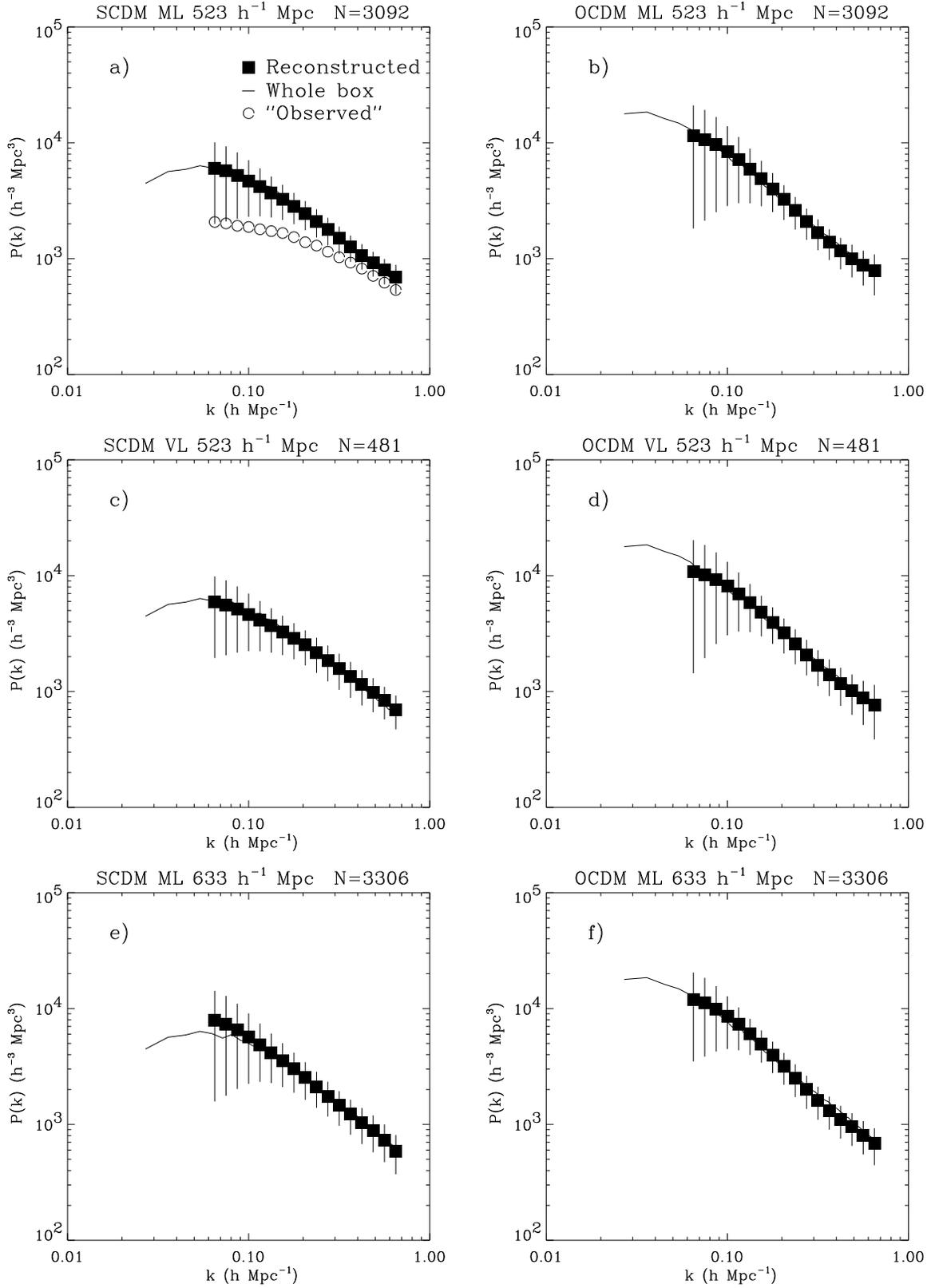}
\caption{Deconvolved power spectra from 50 mock ESP catalogues
extracted from SCDM (left panels) and OCDM (right panels) 
   simulations.  The filled squares and the error bars give the
ensemble average and standard deviation of the 50 mock samples.  The solid
line is the corresponding power spectrum computed from the whole
simulation box using all particles. $N$ denotes the average number of
particles among the mock catalogues. The three pairs of panels, from
top to bottom, refer to the three different kinds of subsamples ESPm523,
ESP523, and ESPm633, as in the case of the real data.  In panel a) we 
plot also the power spectrum estimate before applying the
deconvolution procedure (open circles).
}\label{fig6}
\end{figure*}
On the 
other hand, for small values of $k$, i.e. for spatial wavelengths
comparable to the typical scales of the window (which are quite
small, due to the strongly anisotropic shape), the true 
power is spread over a wide range of wavenumbers.

\section {Numerical Tests}\label{ESP_num}

We test the whole procedure for estimating the power spectrum through
$N$-body simulations that we have run assuming some cosmological models
(Carretti 1999). 
The results of these simulations can be considered as a Universe, from
which we can extract mock catalogues with the same features of the ESP
survey (geometry, galaxy 
density, field completeness, selection function). We then apply to
such mock catalogues the whole power spectrum estimate procedure (convolved
power spectrum estimator and deconvolution tecnique) and we compare
the result with the {\em true} power spectrum obtained from the whole set of
particles. 

The simulations were performed on a Cray T3E at CINECA supercompunting 
center (Bologna)
using a Particle-Mesh (PM) code (Carretti \& Messina 1999) and adopting two
cosmological models: an unbiased Standard Cold Dark
Matter (SCDM: $\Omega_{\rm o} = 1$, $h = 0.5$, $\sigma_8 = 1$) and an unbiased
Open Cold Dark Matter with shape parameter $\Gamma = 0.2$ 
(OCDM: $\Omega_{\rm o} = 0.4$, $h = 0.5$, $\sigma_8 = 1$).
They were run with a box
size of $700 h^{-1}$ Mpc, $512^3$ grid 
points and $512^3$ particles, in order to reproduce a volume which can
contain all catalogues selected for the analysis (max depth $633
h^{-1}$ Mpc) and to select a realistic number of mock galaxies for the
magnitude--limited catalogues.
From each simulation box, we randomly choose a particle as origin and
extract sets of particles with the same features of the three ESP
catalogues.  The magnitude--limited selection is then reproduced by
simply assigning a weight corresponding to the observed selection
function.  

From each simulation and for each ESP subsample we construct 50
independent mock catalogues.  The average number of particles over the
50 realisations is set to the number of galaxies observed in the corresponding
true ESP sample.  The power spectrum estimator is then applied to
each of the 50 mock galaxy catalogues, producing an independent
estimate of $P(k)$ for that specific model and sample geometry.  From
each set of realisations, a mean $P(k)$ and its standard deviation can
finally be computed and compared to the true power spectrum
obtained from all the particles.

A general result from this exercise is that for $k>0.065 h$~Mpc$^{-1}$
the systematic power suppression by the window function convolution is
properly corrected for by our procedure, i.e. we are able to fully recover
the input $P(k)$.  
In figure~\ref{fig6} we show the reconstructed power spectra from
the three sets of mock catalogues, compared to the ``true'' ones, both
for the SCDM and OCDM simulations.  In panel a), in particular, we
also show (open circles) the raw power spectrum before applying the
Lucy deconvolution, to emphasize the dramatic effect of the ESP window
function on all scales.
It is evident that for $k>0.065 h$~Mpc$^{-1}$ the mean deconvolved
power spectra are a very good reconstruction of the original ones.
In particular, in the SCDM case where the spectrum
turnover scale is well sampled by the simulation box, the technique is
able to nicely follow the change of shape at small $k$'s.  This is
important, because guarantees that the deconvolution method has enough
resolution as to follow possible features in the data power spectrum,
while at the same time recovering the correct amplitude.
At smaller $k$'s the error bars explode, and the results become
meaningless.  In the case of the deepest sample, ESPm633, the reconstruction
shows a small systematic overestimate of the amplitude for the SCDM spectrum
on the largest scales, i.e. the reconstruction algorithm seems to have
difficulty in following the curvature of the spectrum accurately.
This is probably due to the rather small value of the selection
function in the most distant part of the sample, which puts too large
a weight 
on the
distant objects.  Rather than indicating a difficulty in the
technique, this is 
probably 
telling us that it is 
safer to
truncate the data at smaller distances, as for 
sample ESPm523. 

Comparing the results from the SCDM and OCDM mock catalogues, 
we have checked
that the fractional errors for the two cases are quite similar.  
Not knowing a priori the correct cosmological model,
rather than choosing one of the two models as representative, we prefer
to average the fractional errors measured from the two models.

\begin{figure}
 \epsfxsize=\hsize \epsfbox{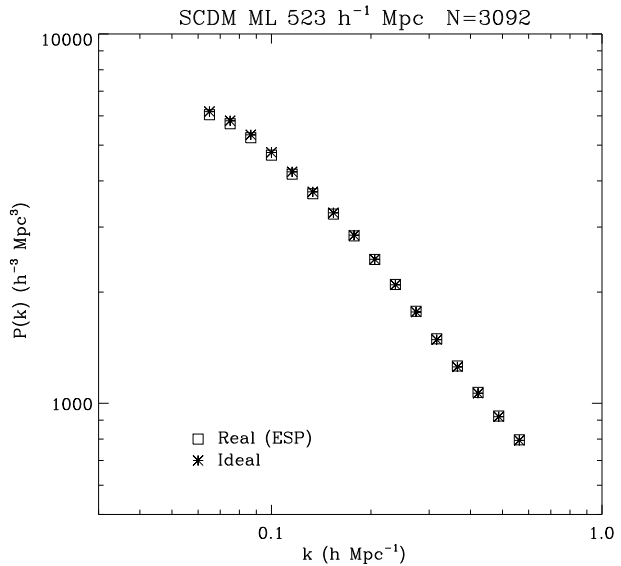} \hfil
 \caption{Test for the effect of redshift incompleteness. The power spectrum
   has been computed for
   50 mock samples extracted from the SCDM simulation,
   both in the ideal case of a full
   redshift coverage of a sample as ESPm523, and in the real situation,
   i.e. including the field-to-field incompleteness and correcting for it
   through the weighting scheme.
   }
 \label{fig7}
\end{figure}

Using the mock catalogues, 
we can also
evaluate directly
the possible effects of the field incompleteness on the power spectrum
estimate.  Figure~\ref{fig7} compares the mean power spectra obtained 
from one set of 50 SCDM mock samples both in the ideal case (all fields
complete) and when the ESP field--to--field incompleteness is
introduced and corrected for.  It is clear that the incompleteness is
correctly taken into account by the weighting scheme.
Error bars (not reported for clarity) are also very similar.

\section {The Power Spectrum of ESP galaxies}\label{ESP_pow}

The numerical tests performed have given us an estimate of the reliability of
our method to reconstruct the true power spectrum, so that we can now apply it
to the three galaxy subsamples ESPm523, ESP523 and ESPm633. 

\begin{figure*}
 \epsfxsize=\hsize \epsfbox{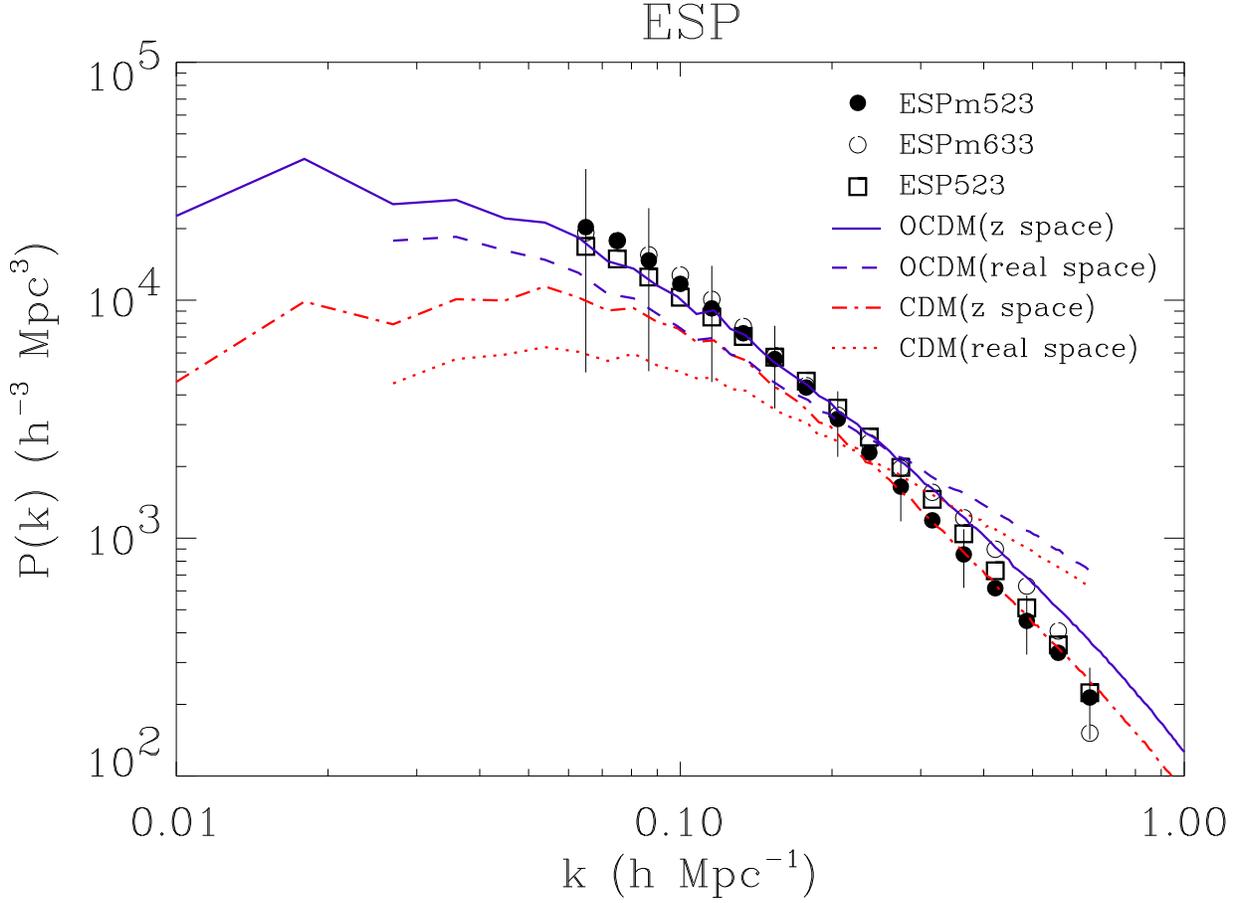} \hfil
 \caption{The final deconvolved estimates of the ESP power spectrum
          from the three samples.  The plot shows also the power spectra
	  computed from the two
	  simulations described in the text both in redshift-- and in the
	  real--space. Note how redshift distorsion effects modify
	  the power spectra increasing the apparent power on large
	  scales and reducing it on small ones.}
 \label{fig8}
\end{figure*}

The final results of the computation are shown in figure~\ref{fig8} and
Table~\ref{tab_value}. 
\begin{table*}
 \centering
  \caption{Results of the power spectrum estimates from the three subsamples of
           the ESP survey.  Errors are estimated from
           50 mock realisations of the samples, as detailed in the text.}
  \begin{tabular}{@{}ccccccc@{}}
     \hline \\
     $k$ ($h$Mpc$^{-1})$ & $P(k)_{ESPm523}$ & $1\sigma_{ESPm523}$ &
     $P(k)_{ESPm633}$ & $1\sigma_{ESPm633}$ & $P(k)_{ESP523}$ &
     $1\sigma_{ESP523}$ \\
     \hline\\
   0.065 & 20284.1 & 15313.7 & 19215.2 & 14497.3 & 16830.1 & 12935.8 \\
   0.075 & 17831.0 & 12781.3 & 17791.6 & 12565.5 & 14907.9 & 10737.9 \\
   0.087 & 14703.1 &  9671.4 & 15515.0 &  9786.9 & 12507.9 &  8122.6 \\
   0.100 & 11714.0 &  6853.1 & 12750.5 &  6921.8 & 10288.4 &  5867.9 \\
   0.115 &  9238.4 &  4711.6 & 10077.0 &  4658.0 &  8496.4 &  4210.6 \\
   0.133 &  7263.2 &  3200.0 &  7747.6 &  3103.9 &  7035.9 &  3024.2 \\
   0.154 &  5657.4 &  2149.3 &  5847.8 &  2109.7 &  5760.2 &  2175.6 \\
   0.178 &  4295.9 &  1437.8 &  4371.9 &  1491.0 &  4568.9 &  1581.1 \\
   0.205 &  3168.1 &   972.4 &  3284.7 &  1092.9 &  3520.3 &  1177.4 \\
   0.237 &  2289.4 &   671.1 &  2514.1 &   827.7 &  2661.9 &   893.5 \\
   0.274 &  1645.7 &   470.0 &  1964.7 &   638.8 &  1983.4 &   674.5 \\
   0.316 &  1187.8 &   333.6 &  1557.9 &   505.0 &  1455.4 &   500.3 \\
   0.365 &   853.9 &   236.0 &  1216.7 &   395.9 &  1042.6 &   360.9 \\
   0.422 &   615.7 &   169.3 &   898.3 &   295.7 &   729.0 &   254.8 \\
   0.487 &   449.0 &   125.1 &   629.0 &   209.3 &   509.1 &   180.1 \\
   0.562 &   329.5 &    95.6 &   406.9 &   137.6 &   355.6 &   128.9 \\
   0.649 &   213.8 &    71.5 &   151.4 &    54.8 &   223.9 &    91.6 \\
     \hline
\end{tabular}
\label{tab_value}
\end{table*}
The error bars are partially reported only for ESPm523
to avoid confusion (the errors are similar for the three samples).
The three estimates of the power spectrum are well consistent
with each other.
Given the large amplitude of the errors ($\sim 30$\% of $P(k)$ for 
 $k>0.15 h$ Mpc$^{-1}$, 50\% for $k\sim 0.1$ and 75\% for $k\sim 0.065$)
the small differences in the slopes are not significant. In general, we can
safely say that the power spectrum of 
ESP galaxies follows a power law $P(k) \propto k^n$ with $n\sim -2.2$
for $k>0.2 h$ Mpc$^{-1}$, and $n\sim -1.6$ for $k < 0.2 h$ Mpc$^{-1}$.
In the range $0.065<k<0.6\,h$ Mpc$^{-1}$ there is no meaningful difference
between the three estimates, which are therefore independent of the catalogue
type (magnitude-- or volume--limited) and of the catalogue depth.

In figure~\ref{fig8} we have also plot, for comparison, redshift--
and real--space power spectra computed from the simulations described
in section~\ref{ESP_num} (SCDM and $\Gamma=0.2$ OCDM).
Note how the former compensate for non linear evolution at large $k$'s in this
way steepening the slope of $P(k)$ over the whole observed range, which makes
the global slope closer to the observed one.
Despite this comparison to models is
deliberately limited, one can safely say that the data points (especially
below $k=0.1$--$0.2\, h$~Mpc$^{-1}$) are in better agreement with
the power spectrum of the $\Gamma = 0.2$ OCDM model.
This model would reproduce this observation without biasing
(the normalisation adopted by the
simulation is $\sigma_8 = 1$).

\section{Consistency between Real and Fourier Space}\label{ESP_corr}

\begin{figure*}
\epsfxsize=0.7\hsize \epsfbox{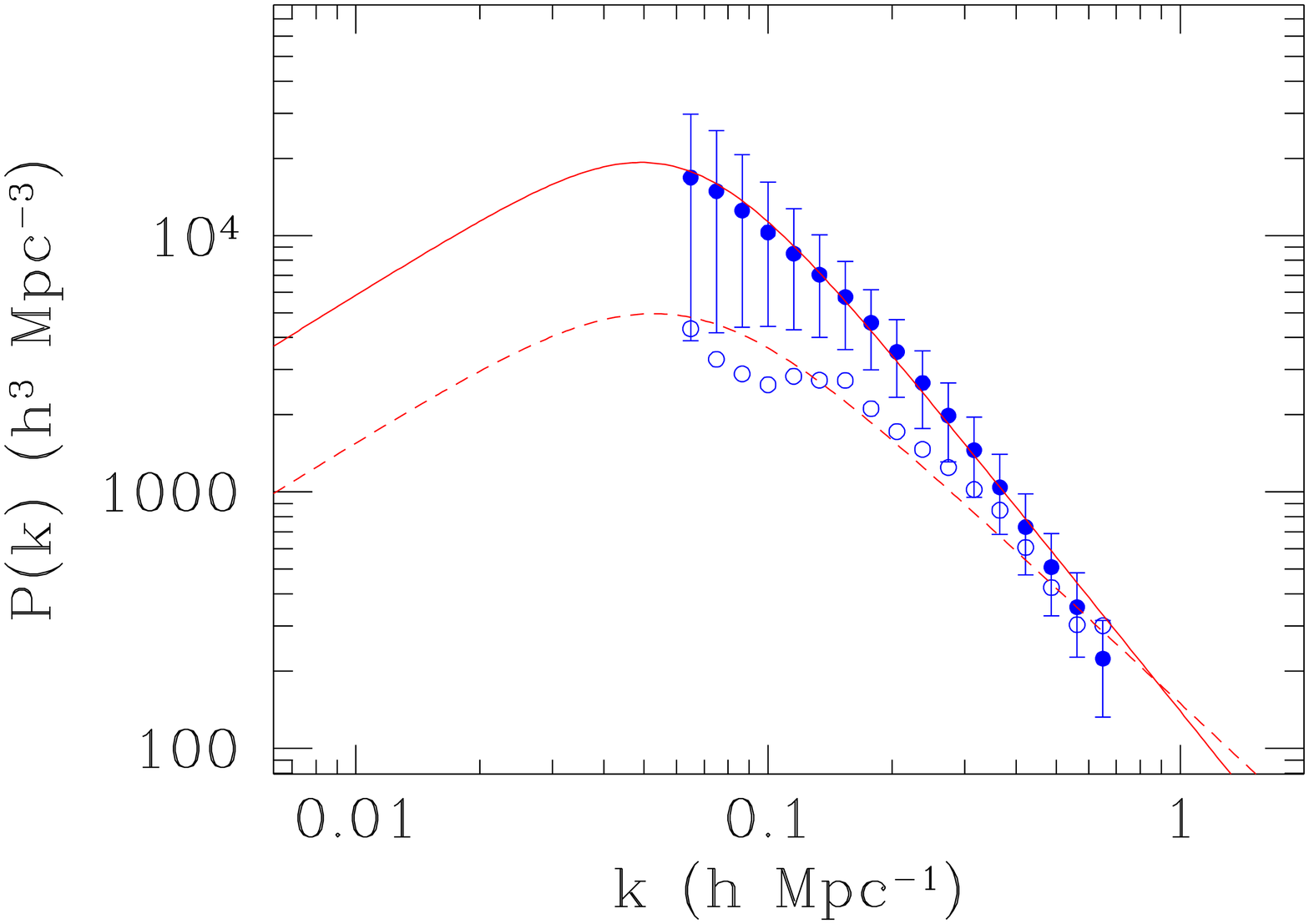} \hfil
 \caption{Fits with a simple phenomenological form of the convolved and
de--convolved P(k) from the ESP523 sample}
 \label{pk_fit}
\end{figure*}
\begin{figure*}
\epsfxsize=0.7\hsize \epsfbox{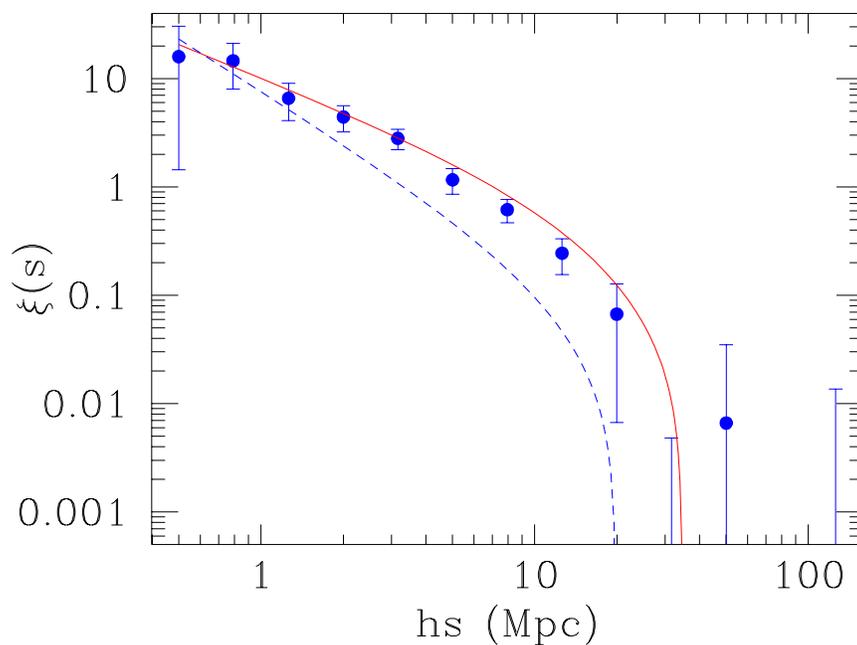} \hfil
\caption{The Fourier transform of the convolved (dashed line) and de--convolved
(solid line) estimates of the power spectrum compared to the two--point
correlation function of the ESP survey (filled circles), estimated for
essentially the same volume--limited subsample (Guzzo et al. 2000).
The Fourier transform
of the de--convolved estimate is in very good agreement with the
direct measure of $\xi(s)$.  This result shows also how the two--point
correlation function -- for which no kind of correction has been
applied in addition to those of standard estimators -- is
substantially insensitive to the effect of the window function.}  
 \label{xi}
\end{figure*}

It is interesting to compare the Fourier transform of the
ESP power spectrum estimated in this work with the two--point
correlation function measured independently from the same sample
(Guzzo et al. 2000).   This exercise is a further check of
the robustness and self--consistency of the estimate of $P(k)$.  In addition,
it is of specific interest to verify the 
effect of the survey geometry/window function in real and Fourier space.
To simplify the procedure, we have first fitted the observed $P(k)$
with a simple
analytical form with two power laws connected by a smooth turnover
(e.g. Peacock 1997)
\begin{equation}
P(k) = {(k/k_o)^\alpha \over 1+\left(k / k_c\right)^{\alpha-n}} \,\,\,\,\, ,
\end{equation}
where $k_o$ is a normalisation factor, $k_c$ gives essentially the
turnover scale, $n$ is the large--scale primordial index (here fixed
to $n=1$), and $\alpha$ gives the slope for $k\gg k_c$.  We have used
this function to 
reproduce the global shape of 
both the convolved and de--convolved estimates of
$P(k)$ 
from the ESP523 sample.  In terms of selection function,
this sample is the closest to one of the volume--limited samples used
by Guzzo et al. (2000) to estimate $\xi(s)$ from the same data.
Figure~\ref{pk_fit} shows how this 
form provides a good description of the 
ESP power spectrum,
with the deconvolved one characterised by
$k_o=0.080\, h$~Mpc$^{-1}$, $k_c=0.062\, h$~Mpc$^{-1}$, $\alpha=-2.2$
(note that while the slope $\alpha$ is a stable value, the turnover
scale $k_c$ is very poorly constrained, given the limited range
covered by the data). 

Figure~\ref{xi} shows the Fourier transform of the two fits, compared
to the direct estimate of $\xi(s)$ by Guzzo et al. (2000).  Two
main comments should be made here.  First, our "best" estimate of $P(k)$,
deconvolved for the ESP window function according to our recipe,
reproduces rather
well the observed two--point correlation function (solid line).  Note how the
Fourier transform of the simple 
direct estimate suffers from a systematic lack of power as a function
of scale (dashed line), as we expected from our results on the mock
samples.  The second, more general comment 
concerns the stability of the two--point correlation function.  One
might naively think that the narrowness of the explored volume, which
gives rise to the window function in Fourier space, should affect in a
similar way also the estimate of clustering by the two--point correlation
function.  Figure~\ref{xi} shows that this is not the case.  In fact, the
points showed here have not been subject to any kind of correction
(Guzzo et al. 2000), a part from those which are standard in the estimation
technique to take into account the survey boundaries. 
Still, they seem to sample clustering to the largest
available scales in a reasonably unbiased way, without basically
being affected by the survey geometry.

\section {Comparison to Other Redshift Surveys}\label{ESP_comp}

\begin{figure*}
  \epsfxsize=0.9\hsize \epsfbox{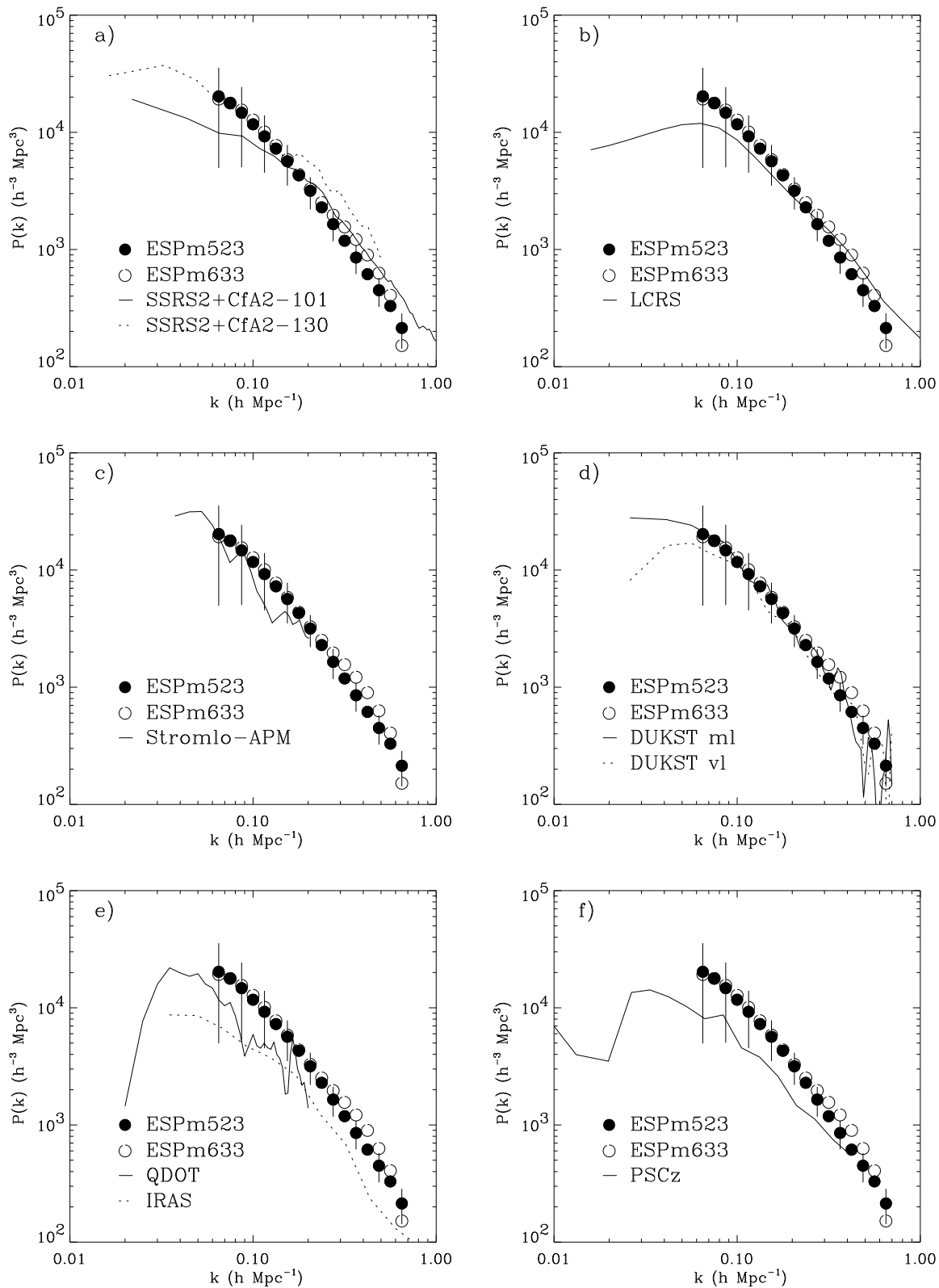}
 \caption{Comparison of the ESP P(k) with results
          from other surveys, as indicated by the labels (see text for
          references).   Error bars for the ESP points are reported only
          partially for clarity.}
 \label{fig11}
\end{figure*}

In the six panels of figure~\ref{fig11}, we compare the power spectrum for the
ESPm523 and ESPm633 samples to a variety of results from previous surveys, both
selected in the optical and infrared (IRAS) bands.  In general, there is a good
level of unanimity among the different surveys concerning the slope of $P(k)$ over
the range sampled by the ESP estimate.  Optically--selected surveys show a good
agreement also in amplitude, with a possible minor 
differential
biasing effect in the case of 
CfA2--SSRS2--130 (panel a, da~Costa et al. 1994), which is a volume--limited 
sample containing galaxies brighter than $\sim M^*-1.5$.  The effect
of different biasing values
is more evident in the comparison 
to IRAS--based surveys
(IRAS 1.2~Jy, Fisher et al. 1993; QDOT, Feldman et al. 1994; PSCz,
Sutherland et al. 1999) in panels e and f. 

Particularly relevant is the comparison to the results of the Durham/UKST
galaxy redshift survey (DUKST, Hoyle et al., 1999) (panel d).  This survey
is selected from the same parent photometric catalogue as
the ESP (the EDSGC) and contains a comparable number of redshifts.
However, it is less deep (bj~17), while covering a much larger solid
angle by measuring redshift in a sparse--sampling fashion, 
picking one galaxy in three.  This produces a window function which is
essentially complementary to that of the ESP survey, with a good
sampling of long wavelengths and a poor description of small--scale
clustering, which on the contrary is well sampled by the ESP 1-in-1
redshift measurements. 
  The agreement between these two data sets is impressive.  This is a
further confirmation of the quality of the deconvolution procedure we have
applied to the ESP data, given the rather 3D shape of the DUKST volume which
makes the window function practically negligible for this survey.  Significantly
more noisy is the estimate from the similarly $b_J$--selected Stromlo-APM
redshift survey (Tadros \& Efstathiou 1996), most probably because of the very
sparse sampling of this survey and the smaller number of galaxies.

Finally, panel b) shows a comparison with the data from the $r$--selected LCRS
(Lin et al. 1996). 
The power spectrum from this survey has a flatter slope with respect to our
estimate from the ESP.  More in general, it is flatter than practically all
other power spectra shown in the figure.   This is somewhat suspicious, as the
two--point correlation functions agree rather well for ESP, LCRS, Stromlo-APM
and DUKST (Guzzo 1999), and might be an indication that the effect of the window
function has not been fully removed from the estimated spectrum. 

\section {Summary and Conclusions}\label{conc}

The main results obtained in this work can be summarised as follows.

\begin{itemize}

\item We have developed a technique to properly describe the
ESP window function analytically, and then deconvolve it from the
measured power spectrum, to obtain an estimate of the galaxy power
spectrum.  The tests performed on a number of mock catalogues
drawn from large $N$--body simulations show that the technique is able
to recover the correct shape of $P(k)$ down to wavenumbers $k\simeq
0.065\, h$ Mpc$^{-1}$.  In general, this technique for describing the
window function analytically can be applied to any redshift survey
composed by circular patches on the sky (e.g. the ongoing 2dF
survey). In addition to its mathematical elegance, it has some
computational advantages over the traditional method for recovering
the survey window function, normally based on the generation of large
Montecarlo poissonian realisations.  

\item The final estimates of the ESP power spectrum, extracted from three
subsamples of the
survey, are in good agreement within the error bars. The bright
volume--limited sample does not show a clear difference in amplitude
with respect to the apparent--magnitude limited ones.  This agrees
with the similar behaviour found for the two--point correlation
function, i.e. a negligible evidence for luminosity segregation even for
limiting absolute magnitudes $M_{b_J}\sim -20$ (Guzzo et al. 2000).
This is only apparently in contrast with the results of Park et
al. (1994), who found evidence for luminosity segregation studying the
amplitude of the power spectrum in the CfA2 survey.  In fact, that
analysis concentrates on a range of luminosities about 1.5
magnitude brighter than $M^*$, which for the CfA2 survey has a value of -18.8
(Marzke et al. 1994), i.e. nearly one magnitude fainter than for 
the ESP.  This also agrees with the results of Benoist et
al. (1996), who studied the correlation function for the SSRS2 sample,
finding negligible signs of luminosity segregation for $M>M^*$.  

\item All three estimates of $P(k)$ show a similar shape, with a well
defined power--law $k^n$ with $n\simeq -2.2$ for $k\ge 0.2\,h$
Mpc$^{-1}$, and a smooth bend to a flatter shape ($n\simeq -1.6$) for
smaller $k$'s. The smallest wavenumber where a meaningful
reconstruction can be performed ($k\simeq 0.065\,h$ Mpc$^{-1}$),
does not allow us to explore the range of scales where other power
spectra seem to show a flattening and hints for a turnover.
In the framework of CDM models, however, the well--sampled steep slope between
0.08 and 0.3~$h$~Mpc$^{-1}$ 
favours a low--$\Gamma$ model ($\Gamma = 0.2$), consistently with the most
recent CMB observation of BOOMERANG/MAXIMA experiments (Jaffe et al. 2000). 

\item We have verified that the two--point correlation function
$\xi(s)$ is much less 
sensitive to the effect of a 
difficult
window function as that of the ESP, than the
power spectrum.
In fact, the measured correlation function (without any
correction), agrees with the Fourier transform of the power spectrum,
only after 
this has been cleaned of the combination by the window function.
This is an instructive
example of how these two quantities, despite being mathematically equivalent,
can be significantly different in their practical estimates and be very
differently affected by the peculiarities of data samples.

\item When compared to previous estimates from other surveys, the ESP power
spectrum is 
virtually indistinguishable from that of the Durham-UKST survey
over the common range of wavenumbers. In particular, between 0.1 and 1 
$h$~Mpc$^{-1}$ our power spectrum has significantly smaller error bars
with respect to the DUKST, by virtue of its superior 
small--scale sampling. The absence of any systematic amplitude
difference between these two surveys -- both selected from the EDSGC
catalogue, but with complementary volume and sampling choices -- 
is an important indirect indication of the quality of the deconvolution
procedure applied here, and also of the accuracy of the two
independent estimates.  In this respect, a combination of the Durham-UKST
and ESP surveys  
possibly provides
the current best 
measure of $P(k)$ for blue--selected galaxies in the full range $\sim
0.03-1 \,h$ Mpc$^{-1}$. 
It will be very interesting to compare these combined
results to the power spectrum of the
forthcoming 2dF redshift survey, which is also selected in the same $b_J$ band 
to virtually the same limiting magnitude than the ESP.
\end{itemize}

\section*{Acknowledgments}
We thank an anonymous referee for suggestions that helped us to improve the
paper. We thank Hume Feldman,
Huan Lin,
and Michael Vogeley for 
providing us with their power spectrum results in electronic form, and 
Stefano Borgani for his COBE normalisation routine. 
LG and EZ thank all their collaborators in the ESP survey team, for their
contribution to the success of the survey.
This work has been partially supported by a CNAA grant.

\bsp

\label{lastpage}

\end{document}